\documentclass[epj,nopacs]{svjour}
\usepackage{amsmath,amssymb,amsfonts,graphicx,cite,multirow,color,algorithmic,algorithm}
\usepackage[utf8]{inputenc}
\usepackage{amsmath}
\usepackage{caption}
\usepackage{subcaption}
\usepackage{hyperref}
\usepackage{xspace}
\usepackage{longtable}
\usepackage{soul}

\graphicspath{{graphics/}}
\makeatletter
\ifx\input@path\@undefined
\def\input@path{{graphics/}}
\else
\g@addto@macro\input@path{{graphics/}}
\fi
\makeatother

\newcommand{\rot}[1]{\rlap{\rotatebox{45}{#1}}\hspace*{1em}}
\makeatletter
\def\instring#1#2{TT\fi\begingroup
  \edef\x{\endgroup\noexpand\in@{#1}{#2}}\x\ifin@}
\makeatother
\newcommand{\makeflag}[3]{%
\if\instring{#1}{#3}{$\checkmark$}\else\if\instring{#2}{#3}{$\bigcirc$}\else{$-$}\fi\fi%
}

\newcommand{\bsmoptionsBLHA}[1]{%
\makeflag{B}{b}{#1} % BLHA
& 
\makeflag{L}{l}{#1} % semi-leptonic decay
&
\makeflag{F}{f}{#1} % VBF process
&
\makeflag{V}{v}{#1} % anomalous Gauge couplings
&
\makeflag{H}{h}{#1} % anomalous Higgs couplings
&
\makeflag{T}{t}{#1} % Two-Higgs model
&
\makeflag{K}{k}{#1} % Kaluza-Klein model
&
\makeflag{S}{s}{#1} % Spin-2 model
&
\makeflag{M}{m}{#1} % MSSM
}

\newcommand{\bsmgfoptionsBLHA}[1]{%
\makeflag{B}{b}{#1} % BLHA
&
\makeflag{G}{g}{#1} % gluon-fusion process
&
\makeflag{L}{l}{#1} % semi-leptonic decay
&
\makeflag{H}{h}{#1} % anomalous Higgs couplings
&
\makeflag{T}{t}{#1} % general 2HDM
&
\makeflag{M}{m}{#1} % MSSM
}

\newcommand{\Vbfnlo}{\textsc{Vbfnlo}}
\newcommand{\Openmpi}{\mbox{\textsc{Openmpi}}}
\newcommand{\Herwig}{\mbox{\textsf{Herwig}}~}
\newcommand{\HerwigNS}{\mbox{\textsf{Herwig}}}
\newcommand{\Matchbox}{\textsc{Matchbox}~}

\newcommand{\GeV}{\ensuremath{\,\mathrm{GeV}}\xspace}
\newcommand{\TeV}{\ensuremath{\,\mathrm{TeV}}\xspace}

\newcommand{\bea}{\begin{eqnarray}}
\newcommand{\eea}{\end{eqnarray}}

\newcommand{\sect}[1]{Section~\ref{#1}}

\newcommand{\appen}[1]{Appendix~\ref{#1}}

\preprint{DESY-24-043\\KA-TP-09-2024}

\title{Release Note -- VBFNLO 3.0}

\author{%
  Julien Baglio\inst{1}, %
  Francisco Campanario\inst{2}, %
  Tinghua Chen\inst{3}, %
  Heiko Dietrich-Siebert\inst{4}, %
  Terrance Figy\inst{5}, %
  Matthias Kerner\inst{4}, %
  Michael Kubocz\inst{6}, %
  Duc Ninh Le\inst{7}, %
  Maximilian L\"oschner\inst{8}, %
  Simon Pl\"atzer\inst{9,10}, %
  Michael Rauch\inst{4}, %
  Ivan Rosario\inst{2}, %
  Robin Roth\inst{4}, %
  Dieter Zeppenfeld\inst{4}
}
\authorrunning{J.\ Baglio et al.}

\institute{QuantumBasel Schorenweg 44B, CH-4144 Arlesheim, Switzerland \and
Theory Division, IFIC, University of Valencia-CSIC, Parque Científico, C/Catedrático José Beltrán, 2, E-46980 Paterna, Spain \and
IT Research Cyberinfrastructure, University of Delaware Newark, DE 19716, USA \and
Institute for Theoretical Physics, Karlsruhe Institute of Technology (KIT), 76128 Karlsruhe, Germany \and
Department of Mathematics, Statistics and Physics, Wichita State University, 1845 Fairmount Street, Wichita, KS\ 67002, USA \and
Institut f\"ur Theoretische Teilchenphysik und Kosmologie, RWTH Aachen University, D52056 Aachen, Germany \and
Phenikaa Institute for Advanced Study, Phenikaa University, Hanoi 12116, Vietnam \and
Deutsches Elektronen-Synchrotron DESY, Notkestr.\ 85, 22607 Hamburg, Germany \and
Institute of Physics, NAWI Graz, University of Graz, Universit\"atsplatz 5, A-8010 Graz, Austria \and
Particle Physics, Faculty of Physics, University of Vienna, Boltzmanngasse 5, 1090 Wien, Austria
}

\date{\today}

\abstract{
  \Vbfnlo{} is a flexible parton level Monte Carlo program for
  the simulation of vector boson fusion~(VBF), QCD-induced single and
  double vector boson production plus two jets, and double and triple
  vector boson production (plus jet) in hadronic collisions at
  next-to-leading order~(NLO) in the strong coupling constant, as well
  as Higgs boson plus two and three jet production via gluon fusion at
  the one-loop level. For the new version -- \textsc{Version 3.0} --
  several major enhancements have been included. An
  interface according to the Binoth Les Houches Accord (BLHA) has been
  added for all VBF and di/tri-boson processes including fully
  leptonic decays.
  For all dimension-8 operators affecting vector boson scattering
  (VBS) processes, a modified T-matrix unitarization procedure has been
  implemented. Several new production processes have been added, namely
  the VBS $Z\gamma jj$ and $\gamma \gamma jj$ processes at NLO, $\gamma \gamma jj $, $WWj$ and
  $ZZj$ production at NLO including the loop-induced gluon-fusion
  contributions and the gluon-fusion one-loop induced $\Phi jjj$ ($\Phi$ is a CP-even or CP-odd scalar boson) process at LO,
  retaining the full top-mass dependence. Finally, the code has been parallelized using \Openmpi.
}

\begin{document}

\maketitle

%% start contents %%%%%%%%%%%%%%%%%%%%%%%%%%%%%%%%%%%%%%%%%%%%%%%%%%%%%%%%%%%%%%
%%%%%%%%%%%%%%%%%%%%%%%%%%%%%%%%%%%%%%%%%%%%%%%%%%%%%%%%%%%%%%%%%%%%%%%%%%%%%%%%

\section{Introduction}
\label{sec:Introduction}
The \textsc{Lhc} has probed the Standard Model~(SM) of particle physics
to an unprecedented level of accuracy. After the discovery of a
narrow scalar resonance, with a mass around $m_H=125 \GeV$, compatible
with the Higgs particle of the SM, and the lack of new heavier
resonances, the beyond SM physics discovery potential of the
\textsc{Lhc} depends heavily on our
ability to provide accurate cross-section predictions for both signal
and background processes of multi-particle production processes involving jets, leptons, photons and missing energy.

A precise description of the hard, multi-particle production processes is needed,
as well as a method for simulating the measurable hadronic final
states. Furthermore, the flexibility to impose kinematic cuts is
mandatory for processes involving QCD radiation, for example to reduce 
backgrounds or to account for the geometry of the detector. This makes
analytic phase-space integration unfeasible and the implementation
of results in the form of Monte Carlo programs becomes the method of
choice.

Reaching these goals requires at least next-to-leading order~(NLO) QCD
calculations presented in the form of parton level Monte Carlo~(MC)
generators, which are an efficient solution when it comes to final
states characterized by a high number of jets and/or identified
particles.

To fully compare with the observed measurements, one needs
parton shower event generators which properly account for
the high final-state multiplicity due to QCD radiation, adding QCD
emissions to fixed-order processes.

\Vbfnlo{} \cite{Arnold:2008rz, Baglio:2011juf, Baglio:2014uba} is a flexible MC program for vector boson fusion
(VBF) and vector boson scattering (VBS),
QCD-induced single and double vector boson production plus two
jets, and double and triple vector boson (plus jet) production
processes at NLO QCD accuracy. Furthermore, the electroweak
corrections to Higgs boson production via VBF (which are of the same
order of magnitude as the QCD corrections in the experimentally
accessible regions of phase-space) have been included.  Since real
emission processes are part of the NLO cross-sections, \Vbfnlo{}
provides the means to calculate cross-sections for the corresponding
process with one additional jet at leading order (LO) in the strong
coupling.

\Vbfnlo{} can be run in the Minimal Supersymmetric Standard Model
(MSSM), with real or complex
parameters, for Higgs boson production via VBF. Anomalous couplings
of the Higgs boson and electroweak gauge
bosons have been implemented for a multitude of
processes. Additionally, two Higgsless extra dimension models are
included -- the Warped Higgsless scenario and a Three-Site Higgsless
Model -- for selected processes. These models can be used to simulate
the production of technicolor-type vector resonances in VBF and triple
vector boson production. Diboson plus two jets production via VBF can
also be run in a spin-2 model~\cite{Frank:2012wh} and in a model with two Higgs
resonances.

In addition, the simulation of CP-even and
CP-odd Higgs boson production in gluon fusion, associated with
up to three additional jets, is implemented at LO QCD.
For these gluon fusion processes, the full top- and
bottom-quark mass dependence of the one-loop contributions in the
Standard Model (SM), in the MSSM and in a generic two-Higgs-doublet model is included.

Arbitrary cuts can be specified as well as various scale choices. Any
currently available parton distribution function~(PDF) set can be used
through the \textsc{Lhapdf} library. 
In addition, 
a selected set of PDFs are hard-wired into the code (e.g. CT18~\cite{Hou:2019efy}). At leading
order the program is capable of generating event files in the Les
Houches Accord~(LHA) and the \textsc{HepMC} format, while for next-to-leading-order
this can be done through the Binoth Les Houches Accord (BLHA) interface \cite{Binoth:2010xt,Alioli:2013nda}, 
for the processes where it is 
implemented. When working in the MSSM,
the SUSY parameters can be input via a standard SLHA file.

In this article, we present the new version of the program
\Vbfnlo{}, which includes as a major feature the creation of an
interface at NLO according to the BLHA for
all VBF and di/tri-boson processes including fully leptonic decays,
which allows performing full simulations down to the particle level,
if interfaced with a MC event generator, which will take care of
parton shower and hadronization effects.

The \Vbfnlo{} program has been released via a series of versions 2.0, 2.5.0, 2.6.0, and 2.7.0.~\cite{Arnold:2008rz, Baglio:2011juf}.
This version supersedes version 2.7.0, with the following
new features:
\begin{itemize}
\item  A BLHA interface for all VBS, double and triple vector boson production processes.
\item  Parallelization of the code with \Openmpi.
\item The dimension~8 operator $\mathcal{O}_{S,2}$ has been
       added.
     \item For VBS, the K-matrix unitarization procedure has been implemented
       for the dimension~8 operator $\mathcal{O}_{S,1}$ and the
       isospin-conserving combination
       $\mathcal{O}_{S,0}\equiv\mathcal{O}_{S,2}$. For general dimension~8
       operators, the so-called  T$_u$ model is implemented as an
       alternative unitarization procedure.
     \item Linking with \textsc{Lhapdf} v6 has been enabled.
     \item The following new processes have been added: \\ $WWj$ and
       $ZZj$ production including the loop-induced gluon-fusion~(GF)
       contributions of $\mathcal{O}(\alpha_s^3)$.  QCD-induced
       $ZZjj$, VBS and QCD-induced $Z\gamma jj$ and $\gamma \gamma
       jj$, $Z(\rightarrow \ell^+ \ell^-)Z(\rightarrow \nu
       \bar{\nu})\gamma$ production. The GF one-loop induced $\Phi jjj$
       process at LO, with $\Phi \in (h,A)$, with $h$ and $A$ general
       CP-even and CP-odd Higgs particles.
     \item Higgs decays have been implemented for the VBF-HHjj
       process.
\end{itemize}
Additionally, several bugfixes and smaller improvements have also been included. The
complete list can be found in the {\tt CHANGELOG.md} file included in the
tarball.

\subsection{Availability}
The new program version, together with other useful tools and
information, can be obtained from the webpage --
\url{https://ific.uv.es/vbfnlo/}.

In order to improve our response to user queries, all problems and
requests for user support should be reported via email to 
\href{mailto:vbfnlo@ific.uv.es}{vbfnlo@ific.uv.es}. \Vbfnlo{} is
released under the GNU General Public License (GPL) version 2 and the
MCnet guidelines for the distribution and usage of event generator
software in an academic setting, which are distributed together with
the source, and can also be obtained from
\url{http://www.montecarlonet.org}.

\subsection{Documentation and further details} 
The \Vbfnlo{} webpage -- \url{https://ific.uv.es/vbfnlo/} -- contains,
in addition to the updated manual, the previous versions of the
program.  Relevant publications and theses are also listed.  Useful
tools, such as for form factor calculation and for combining multiple
independent runs (with different random seeds), are available under
{\tt Downloads}.  A complete list of processes is provided there, also available in
\appen{app:proc_list}.

\subsection{Prerequisites} 
The basic installation requires GNU {\tt make}, a 
{\sc Fortran}~95\footnote{{\tt gfortran} and {\tt ifort} have been tested. Pure
FORTRAN77 compilers like {\tt g77} are no longer supported.} and a C++
compiler. \Vbfnlo{} offers the possibility of using the
\textsc{Lhapdf}\footnote{\tt
\url{https://lhapdf.hepforge.org/}}~\cite{Whalley:2005nh} library
(versions 5 and 6) for parton distribution functions. In order to
include the electroweak corrections to VBF Higgs production, the
program \textsc{LoopTools}\footnote{\tt
\url{http://www.feynarts.de/looptools/}}~\cite{Hahn:1998yk,Hahn:2006qw}
is required.  Additionally, \textsc{FeynHiggs}\footnote{\tt
\url{http://www.feynhiggs.de/}}
~\cite{Frank:2006yh,Degrassi:2002fi,Heinemeyer:1998np,Heinemeyer:1998yj}
can be linked to the code in order to calculate the Higgs boson sector
of the MSSM, although a SLHA file can be used as an alternative.  If
the simulation of Kaluza-Klein resonances is enabled, an installation
of the GNU Scientific Library~(GSL)\footnote{\tt
\url{https://www.gnu.org/software/gsl/}} is required.  \Vbfnlo{}
can also be linked to \textsc{Root}\footnote{\tt
\url{https://root.cern.ch/}} and \textsc{HepMC}\footnote{\tt
\url{http://lcgapp.cern.ch/project/simu/HepMC/}} to produce histograms
and event files in those formats.

\subsection{Installation and running the program}
For detailed installation instructions and available compilation options, please refer to the manual contained in the {\tt doc/} folder of the \Vbfnlo{} distribution.
The installation is performed in a standard \textsc{Unix}-layout, i.e.\ the directory
specified with the {\tt -{}-prefix} option of the {\tt configure} script contains the
following directories:
\begin{itemize}
\item {\tt bin/:} {\tt vbfnlo} executable.
\item {\tt include/VBFNLO/:} \Vbfnlo{} header files. 
\item {\tt lib/VBFNLO/:} \Vbfnlo{} modules as dynamically loadable
  libraries. These can also
be used independently from one of the main programs.
\item {\tt share/VBFNLO/:} Input files and internal PDF tables.
\end{itemize}

The {\tt vbfnlo} executable contained in the {\tt bin}
directory of the installation path looks for the input files in the current working
directory. Alternatively, the path to the input files can be specified explicitly by
passing the {\tt -{}-input} argument to the program, e.g.
\begin{center}
{\tt ./bin/vbfnlo -{}-input=[path to input files]}
\end{center}
when running \Vbfnlo{} from the installation ({\tt prefix}) directory.
The input files contained in the {\tt share/VBFNLO} directory are meant to
represent default settings and should not be changed. We therefore recommend
that the user copies the input files to a separate directory. Here, special
settings may be chosen in the input files and the program can be run from that
directory without specifying further options.  

\Vbfnlo{} outputs a running `log' to the terminal, containing information
about the settings used and integrated cross-sections. In addition, a file (named {\tt xsection.out}) is produced,
which contains only the LO and NLO cross-sections, with the associated statistical errors. 
Additionally, histogram and event files can be output in various forms.

To enable a simple installation test, \Vbfnlo{} has a complete set of
example results, together with input files, in the {\tt regress}
directory, which can be executed typing {\tt make check}. 

\subsection{MPI, parallel jobs and optimised grids}
\label{sec:MPI}
MPI allows to run the \Vbfnlo{} code in parallel. Integration as well as
grid and histogram generation are available with MPI parallelization,
while for event output only single-core runs are supported.

To use it, first load \Openmpi~in your environment (e.g. {\tt module
  load openmpi}), then configure with {\tt -{}-enable-MPI} and the
OpenMPI wrapper {\tt FC=mpifort}.

To run the resulting binary start
\begin{center}
	{\tt mpirun -np 8 /prefix/bin/vbfnlo}
\end{center}
replacing 8 with the number of requested
parallel runs.

For single-core usage MPI and non-MPI runs should agree up to machine
precision, while for multi-core runs, numerical differences up to the
stated statistical uncertainty are expected.

Use the xoroshiro random number generator ({\tt RTYPE=3} in {\tt
  random.dat}) for best performance with MPI. For details see the
manual, Sec.~5.9.

Owing to the complexity of the calculations involved, some of the
processes implemented in \Vbfnlo{} (in particular the spin-2,
triboson plus jet and QCD-induced diboson plus two jets processes)
require a significant amount of time in order to obtain reasonable
results.  There are, however, methods which can be used in order to
reduce the necessary run time or to further improve the statistics.

By using an optimised grid, the number of iterations needed in order
to improve the efficiency of the MC integration can be reduced. This
normally leads to a halving of the CPU time needed to
achieve the same numerical accuracy.
The optimised grids previously
provided on the \Vbfnlo{} webpage are now no longer sustained, 
as they were tailored for a specific set of
kinematic cuts and input parameters. They can be easily generated by the user (see the manual for instructions). A copy of the old set of available grids can still be obtained in the Archive folder of the webpage.

Another method of improving the run time is to run several
statistically independent jobs in parallel with or without MPI and then
combine the results. In order to do this, several input directories
need to be set up containing all the necessary {\tt .dat} input files
for the process.  The variable {\tt SEED} in {\tt random.dat}
(see the manual, Sec.~5.9 ) needs to be set to a different integer value
in each directory.  A short example of the results of a parallel run,
together with their combination, is provided in the regress directory
{\tt regress/100\_Hjj\_parallel}.  On the \Vbfnlo{} website,
there is a shell script which can be used to combine the cross
sections and histograms from parallel runs.

\section{BLHA interface and usage with Herwig 7}
\label{sec:BLHA}
An interface following the BLHA 
\cite{Binoth:2010xt} and its update \cite{Alioli:2013nda} (BLHA2) has
been implemented in \Vbfnlo{}, and extensively tested with Herwig 7, 
for all di-boson, tri-boson and VBS
processes with fully leptonic decays, the semi-leptonic decays are not
yet included. This allows \Herwig to use \Vbfnlo{}\ as a One-Loop Provider
(OLP) through the Matchbox module \cite{Platzer:2011bc}. 
The implemented interface only departs from the BLHA and BLHA2
in that it allows the use of the \Vbfnlo{} phase-space generator, which should
increase the performance noticeably.  Furthermore, instead of using SM
matrix elements, \Vbfnlo{} can provide \HerwigNS\ with matrix elements
calculated using dim-6 and dim-8 operators in the Standard Model Effective Field Theory (SMEFT).

\Herwig is able to perform NLO matching and merging using the matrix
elements provided by \Vbfnlo{}.  The \Matchbox module
\cite{Platzer:2011bc} of \Herwig includes two different matching
procedures: a subtractive (MC@NLO-type) and a multiplicative
(Powheg-type) NLO matching.  Moreover, the implemented processes can
be used for the (N)LO multi-jet merging outlined in Refs.~\cite{Platzer:2012bs, Bellm:2017ktr}.

If \Herwig is installed using the {\tt herwig-bootstrap} script, it
will download and compile a copy of \Vbfnlo{}.  The user can add the
flag {\tt -{}-with-VBFNLO=PATH} during the run of the {\tt
  herwig-boostrap} to use a local build of \Vbfnlo{} instead, 
where {\tt PATH} is the path to the installation directory of the 
local \Vbfnlo{}.

If the user decides to use their own copy of \Vbfnlo{} and it is using
a version of \Herwig 7.3 or higher, they also have to add the
{\tt -{}-enable-custom-blha} flag during the \Vbfnlo{} configuration.
This modifies the standard BLHA interface to allow \Herwig to call
\Vbfnlo{} together with another OLP without name conflicts.
\Herwig 7.3 assumes this naming convention for
\Vbfnlo{} 3.0 or higher. Trying to compile \Herwig 7.3 
with a \Vbfnlo{} 3.0 without this flag will result in a 
compilation error.

Curated input cards have been provided for the user in {\tt
  VBFNLO/regress/runs/BLHA}.  The user is advised to copy this input
card as a starting template and modify it appropriately. Information
about the different options can be found in the \Herwig 
tutorial\footnote{https://herwig.hepforge.org/tutorials/index.html} and
in the \Vbfnlo{} manual.

\section{Unitarization and dimension-8 operators}
\label{sec:unitarization}

Anomalous couplings can be described within an effective field theory (EFT)
framework with operators of higher energy dimension ($d=6$, $d=8$, $\dots$)
\begin{align}
\mathcal{L}_{\text{EFT}}= \sum_i \frac{c_i^{(6)}}{\Lambda^2}\mathcal{O}_i^{(6)}+\sum_i \frac{c_i^{(8)}}{\Lambda^4}\mathcal{O}_i^{(8)}+ ... \, . \label{eq:L_EFT}
\end{align}
Three new dimension-8 EFT operators have been implemented. The complete set can be found in the operator list of the \Vbfnlo{} manual. 
Only bosonic operators are available in our program. 

These higher dimensional operators can produce a violation of unitarity for
some regions of phase-space, in particular for large center-of-mass energies of the
scattering partons. This can be cured by using one of the unitarization procedures
implemented in \Vbfnlo{} (form factors, K-matrix, and T$_u$-model, the latter two being
available for VBS only).
We highly suggest using the T$_u$-model for studying dimension-8 operators in VBS
as it comes without the need for any additional ad hoc input parameters, which is an
improvement over the form-factor method. Moreover, it not only cures unitarity violations
at large center-of-mass energies of the scattering weak bosons, but also for
large virtualities of the incoming vector bosons in VBS by suppressing the
anomalous contribution sufficiently, which represents a major improvement
over the K-matrix unitarization. It is also the only unitarization procedure
in \Vbfnlo{} which works for the full set of dimension-8 operators that are implemented.
For further details on the T$_u$-model and its implementation see Ref.~\cite{Perez:2018kav}.

\section{New processes}
\subsection{New VBS and QCD-induced processes}
\label{sec:VBSandQCD}
Compared to the previous release note \cite{Baglio:2014uba}, several new VBS and QCD-induced processes have been implemented. 
The QCD-induced production of $ZZ jj$  has been calculated in Ref.~\cite{Campanario:2014ioa}, with the two $Z$ boson system decaying either to four charged leptons, $ZZ\rightarrow l^+_1 l^-_1\,l^+_2 l^-_2$, or a pair of charged letons and a pair of neutrinos,  $ZZ\rightarrow l^+ l^-\,\nu\bar \nu$.
For the $Z\gamma jj$ channel, both the VBS as well as the QCD induced production have been implemented in Refs.~\cite{Campanario:2017ffz,Campanario:2014wga} respectively, 
with the $Z$ boson decaying to either a charged lepton pair or a pair of neutrinos.
Similarly, the $\gamma\gamma jj$ production in the VBS and QCD channel have been implemented in Ref.~\cite{Campanario:2020xaf}.
Furthermore, the QCD-induced production of $Z jj$ and $\gamma jj$ have been added to the program. The corresponding VBS production mechanisms have already been available in previous versions of \Vbfnlo{}.

The implementation of all processes follows a similar procedure. First, decay currents for the decays of the vector bosons $V\rightarrow l\bar l$ as well as $V\rightarrow l\bar l\,l'\bar l'$ are constructed, where the latter decay channel contains non-resonant contributions in the production of two vector bosons.
For the VBS processes, in addition the leptonic tensors $V_1^\mu V_2^\nu \rightarrow l\bar l$ and $V_1^\mu V_2^\nu \rightarrow l\bar l\, l'\bar l'$ are constructed, which encode the scattering of the vector bosons $V_i$ in the $t$- or $u$-channel.
Here, $l\bar l$ and $l'\bar l'$ can either represent a charged lepton pair or a pair of neutrinos.  
For the $Z \gamma jj$ production processes, the code was adapted from the $ZZjj$ routine replacing 
the $l'\bar l'$ pair by the final-state photon. This procedure can be repeated to obtain the $\gamma\gamma jj$ code from the $Z \gamma jj$ one. 
The amplitudes of all partonic channels can then be obtained by contracting these currents with the quark lines involved in the process.
Virtual corrections are obtained using the methods described in Ref.~\cite{Campanario:2011cs}, grouping Feynman diagrams which only differ by permutations of the external currents into building blocks, leading to an efficient evaluation of these contributions. 
Gauge-invariance checks, where a current $V^\mu$ is replaced by its momentum, are used to guarantee the numerical stability of the code. 
A rescue system using quadruple precision is switched on when the gauge checks fail.
This means that quadruple precision is used only for a fraction of statistics and, hence,
does not appreciably affect the speed of the program. 
The virtual and real corrections are combined using the Catani-Seymour subtraction algorithm~\cite{Catani:1996vz} to obtain infrared-finite NLO cross-sections. 
The final-state photon is treated using Frixione's smooth cone algorithm \cite{Frixione:1998jh}. 

For all the new VBS $Z \gamma jj$ and $\gamma\gamma jj$ processes, anomalous gauge couplings are implemented 
using the effective Lagrangian with dimension-6 and 8 bosonic operators, see \sect{sec:unitarization} for further 
details about these operators and available unitarization methods. 

Extensive checks have been performed by comparing the tree-level and virtual amplitudes with the ones obtained 
by using automated tools such as Madgraph \cite{Alwall:2007st}, FeynArts-3.4 \cite{Hahn:2000kx}, FormCalc-6.2 \cite{Hahn:1998yk}. 
For the anomalous coupling part, we have cross-checked our implementation at the LO-amplitude level against Madgraph with the FeynRules \cite{Christensen:2008py,Christensen:2009jx} model file \texttt{EWdim6} \cite{Degrande:2013rea,Degrande:2012wf} 
(for dimension-6) and with the FeynRules model files for quartic-gauge couplings \cite{Eboli:2006wa} (for dimension-8). 
Agreement at the machine-precision level has been found at random phase-space points.   
\subsection{Double vector boson production in association with a hadronic jet}
The diboson plus jet processes $W^+W^-j$ and $ZZj$ have been included
at the NLO QCD level. As processes involving multiple electroweak
bosons and jets, these are important channels in which to compare
experimental data with the predictions of the Standard Model. The NLO
QCD corrections to the total cross-sections are sizeable, and have a
non-trivial phasespace dependence. As usual in the \Vbfnlo{} code, the
leptonic decays of the vector bosons are included with all off-shell
and spin correlation effects. These processes were included first as
contributions to the approximate NNLO predictions of the $WW$ and
$ZZ$ production processes presented in
Refs.~\cite{Campanario:2013wta,Campanario:2015nha}. The technical
implementation of the processes follows closely the procedure
described in the previous sub-section. The implementation of the code
has been tested at the amplitude and integrated cross-section level
against Madgraph \cite{Alwall:2007st} for the LO and NLO-real corrections. As for the
virtual corrections, internal checks based on factorization, gauge and
reparametrization invariance, as described in
Ref.~\cite{Campanario:2011cs}, were implemented.
\subsection{New gluon fusion induced processes}
New Gluon Fusion (GF) processes have been implemented in comparison to the previous
release note~\cite{Baglio:2014uba}. The one-loop induced $WWj$ and
$ZZj$ gluon-fusion production processes have been included at LO. The
$ZZj$ channel was first computed for on-shell $Z$'s in
Ref.~\cite{Campanario:2012bh}, where an extensive study of the
stability of the code due to the presence of the Gram determinants was
performed. A rescue system was implemented, using both standard tensor
integrals reduction methods with quadruple precision and dedicated
routines for small Gram determinants. This combination helped to
reduce the impact of quadruple precision routines while minimizing
numerical instabilities to an insignificant level.

The amplitudes are computed by building up a set of four master
Feynman diagrams and attaching to them the corresponding particles, as
described in detail in Ref.~\cite{Campanario:2012bh}. Afterwards,
following the strategy presented in Section~\ref{sec:VBSandQCD}, the
leptonic decays of the vector bosons, including all off-shell and spin
correlation effects, were included. Both processes $WWj$ and $ZZj$
were also included as contributions to the higher order QCD
corrections, ${\cal O} (\alpha_s^3)$, to the predictions of the $WW$
and $ZZ$ production processes presented in
Refs.~\cite{Campanario:2013wta,Campanario:2015nha}.

The full GF one-loop induced $\Phi jjj$ process at LO has been
included, with $\Phi \in (h,A)$, where $h$ and $A$ are general CP-even
and CP-odd Higgs particles, extending the results presented in
Refs.~\cite{Campanario:2013mga,Campanario:2014oua}, which only
contained the purely gluonic channel.
Finite bottom-mass loops have also been included since these
corrections dominate in some BSM scenarios. Similar to the $GFVVj$
processes, the full amplitude is formed from a reduced set of Feynman
master integrals, attaching to each of them the corresponding
particles or currents to build a contributing Feynman diagram. The
amplitudes have been cross-checked internally in the
heavy-top limit, reaching a relative precision at the amplitude level
of $10^{-5}$ or below
for values of the top mass of 50\,$\text{TeV}$.

Due to the numerically challenging one-loop times one-loop hexagon
diagrams, a rescue system based on quadruple precision has been
designed. Based on Ward identities to identify the instabilities, once
an unstable diagram is identified, the scalar and tensor integrals are
evaluated again in quadruple precision. If the point after this procedure
is identified as stable, it is kept; otherwise, the complete amplitude
is set to zero for the given phase space point. This procedure reduces
the instabilities well below the per mille level while only mildly
affecting the CPU time of the code~\footnote{This process is disabled
by default and must be enabled at compilation using the configure
option --enable-processes=all or --enable-processes=ggf.}.

\subsection{Higgs pair production in vector boson fusion including Higgs boson decays}
Higgs pair production plus two jets via VBF is the second largest
production channel at the LHC~\cite{DiMicco:2019ngk}.  At leading
order it is produced by electroweak quark-quark scattering processes,
$qq'\to qq'HH$ and related crossings. The NLO QCD corrections in the
VBF approximation are DIS-type QCD corrections to the scattering quark
lines. The interference with the double
Higgs-strahlung process $qq'\to V^* \to qq' HH$ is negligible, the
latter process being viewed as an entirely separated production
channel, see for example in Ref.~\cite{Baglio:2012np}.

The NLO QCD corrections correspond to virtual corrections to the
$qq'V$ vertex as well as real emission contributions given by a gluon
attached to a quark line, where either the gluon, the quark, or the
anti-quark enter as the initial state particle. They were
implemented following the strategy
described in Section~\ref{sec:VBSandQCD}.

The decay of the Higgs boson pair in the final state can now also be
included in \Vbfnlo{}. More specifically, given that the Higgs decay
width is small, the narrow-width approximation provides accurate
predictions for VBF production. The largest branching ratio in the SM
is for $H\to b\bar{b}$. We have implemented the decay modes $p p\to q q'
HH \to q q' b\bar{b} \gamma\gamma$ and $p p\to q q' HH \to q q'
b\bar{b} \tau^+\tau^-$. The calculation uses the full $2\to 6$
phase-space to define the kinematical quantities. Note that the
calculation can also be interfaced with \Herwig as this is a VBF
production mode, see Section \ref{sec:BLHA} on how to interface
\Vbfnlo{} and \HerwigNS.
%--------------------------------------------------------------------------------
%================= Section ======================================================
%--------------------------------------------------------------------------------

\section{Other changes}

The release \textsc{Version 3.0} includes some changes that alter previous results:

\subsection{NLO calculation of VBF-$Hjjj$}

Bugs have been removed in the virtual and real-emission parts of VBF-$Hjjj$
production, which leads to a decrease of the NLO cross-section of roughly 10\%.

\subsection{NLO calculation of VBF-$Z\gamma jj$ with anomalous couplings}

A bug has been fixed in the real-emission part of VBF-$Z\gamma jj$
production when anomalous couplings were switched on.

\subsection{NLO calculations of $W^-jj$, $W^-Zjj$, $W^-\gamma jj$ production}

A bug in the PDF convolution of the QCD-induced production processes $W^-jj$,
$W^-Zjj$ and $W^-\gamma jj$ has been fixed. The size of the effect,
which can be observed both in the real-emission part of the NLO
calculation and the corresponding LO+jet processes, can reach several
percent, depending on the parameter settings.

\subsection{VBF cuts for $HHjj$}

VBF cuts now also apply to VBF production of a Higgs pair.

\subsection{K-matrix unitarization}

Following the introduction of the dimension~8 operator $\mathcal{O}_{S,2}$, the
K-matrix unitarization procedure for $\mathcal{O}_{S,0}$ has been reworked.
The relation to the parameters $\alpha_4$ and $\alpha_5$ of the electroweak
chiral Lagrangian is now process-universal. To account for the
isospin-conservation of these operators, which is used in the K-matrix
procedure, only the combination 
$c_{S,0}=c_{S,2}$ can
be unitarised, as well as $\mathcal{O}_{S,1}$. For other combinations the T$_u$-model
can be used (see \sect{sec:unitarization}).

\subsection{Calculation of histogram-bin errors}

A bug has been found in the calculation of the statistical errors of histogram bins, 
which leads to an increase of the errors. The mean value of the cross-section (bin-wise) is unchanged. 
The statistical error of the integrated cross-section was correctly calculated, hence not affected by this change. 

\section{Example results}

For illustration, we exhibit a small selection of differential distributions for
the process $pp \rightarrow e^{-} \bar{\nu}_e j j + X$ through VBF at NLO QCD. 
These were computed in two different ways: using \Vbfnlo{} stand-alone
and the \Vbfnlo{}-\HerwigNS\ combination (denoted as \HerwigNS\ here for short). 
Two different runs were perfomed using \HerwigNS, namely a fixed order computation (\HerwigNS\ FO), to check agreement 
with \Vbfnlo{}, and a parton shower matched with \Vbfnlo{}'s amplitude (\HerwigNS\ PS).

The runs are created at a center-of-mass energy of $\sqrt{\rm{s}} = 13 \TeV$ in the 4-flavour scheme 
using fixed scales $\mu_F = \mu_R = 100 \GeV$, and with the MMHT2014 PDFs \cite{Harland-Lang:2014zoa}. 
The following cuts are implemented: charged leptons are required to have
$$
|y_{e}| < 2.5\, ,\quad p_{T, e} > 20 \GeV,
$$
while the jets are defined using the anti-$k_T$ algorithm \cite{Cacciari:2008gp} with the radius parameter $R=0.4$, and
$$
|y_j| < 4.5\, ,\quad p_{T, j} > 30 \GeV.
$$
VBF cuts are also implemented with
$$
\Delta y_{j_1j_2} > 3.6\, , \quad m_{j_1j_2} > 600 \GeV,
$$
where $j_1$ and $j_2$ correspond to the two leading jets in $p_T$.

The resulting integrated cross-sections for the \Vbfnlo{} stand-alone run
(labeled by \Vbfnlo{}) and for the \HerwigNS\ FO run using \Vbfnlo{} as an OLP through
the BLHA interface are
\begin{align}
\sigma_{\Vbfnlo{}} &= 484.18 \pm 0.03 \, \rm{fb} \\
\sigma_{\HerwigNS\ \rm{FO}} &= 483.99 \pm 0.04 \, \rm{fb}. 
\label{Wjj-XS}
\end{align}
The \HerwigNS\ FO integrated cross-section was obtained from the standard output
of the fixed-order run.

\begin{figure}[h!]
\includegraphics[width=0.45\textwidth]{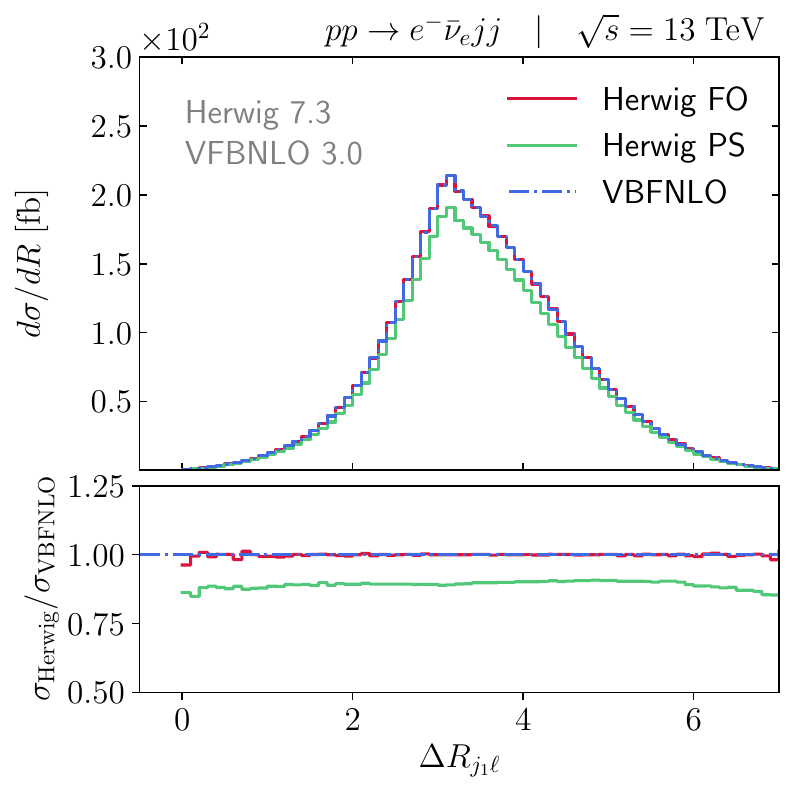}
\caption{$R$-separation between the hardest jet and the electron.}
\label{fig:delta-r-jet1-e}
\end{figure}

\begin{figure}[h!]
\includegraphics[width=0.45\textwidth]{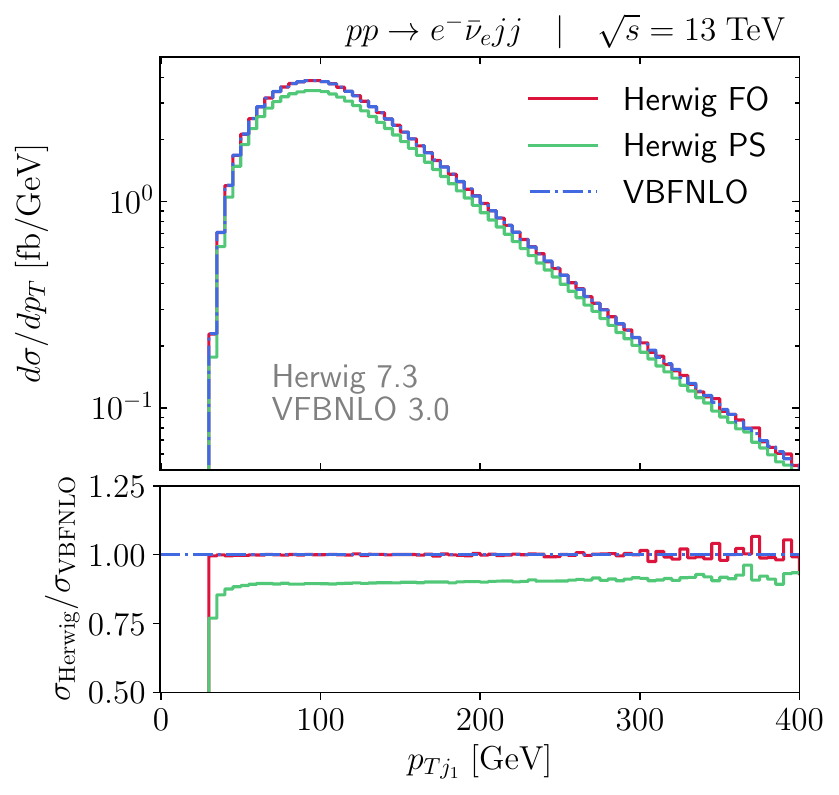}
\caption{Transverse momentum distribution of the hardest jet.}
\label{fig:jet1-pt}
\end{figure}

\begin{figure}[h!]
\includegraphics[width=0.45\textwidth]{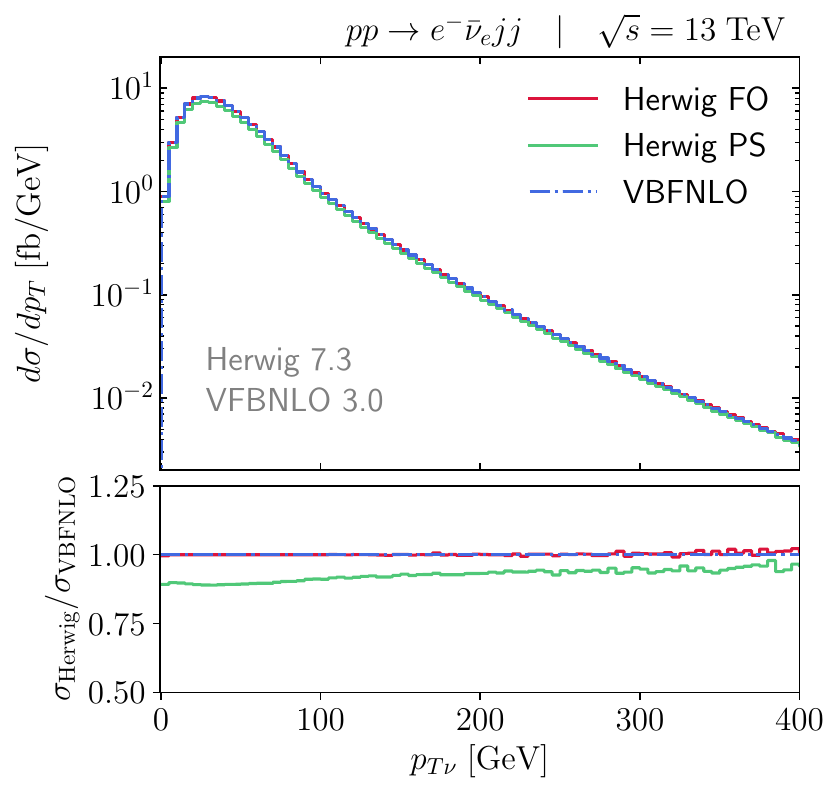}
\caption{Transverse momentum distribution of the neutrino.}
\label{fig:nu-pt}
\end{figure}

In Figs.~\ref{fig:delta-r-jet1-e}-\ref{fig:nu-pt}, we display a few examples of
differential distributions for this process together with the ratios between \Herwig and \Vbfnlo{}.

As an illustration of the new capabilities, we also show the differential
distributions for the \HerwigNS\ PS results. The integrated cross section
with parton shower is about $15\%$ lower than the fixed-order result. This
is due to known migration effects which occur when generation-level cuts
are not sufficiently inclusive~\cite{Rauch:2016pai, Rauch:2016jxo}. For
the $e^{-} \bar{\nu}_e j j$ results shown, we have selected dedicated VBF
cuts at the generation level tailored to improve the efficiency in the
comparison between the \Vbfnlo{} and \HerwigNS\ FO results. 

\section{Summary and outlook}
\label{sec:Summary}

\Vbfnlo{} is a fully flexible partonic Monte Carlo program for vector
boson fusion and scattering, as well as double and triple vector boson production processes at
NLO QCD accuracy. The simulation of the one-loop CP-even and CP-odd
Higgs boson production in gluon fusion, associated with two and three
additional jets, is implemented at leading order, retaining the full top and bottom quark 
mass dependence.

In this release, two major features extend the usability of the
program. First, a BLHA interface for all VBS, double and triple vector
boson productions processes has been created and tested with the event
generator \Herwig and, second, the code has been parallelized using
\Openmpi. Furthermore, additional features like new processes or unitarization
procedures for dimension-8 operators in VBS processes, have been included.

Future improvements are directed along two main lines of development:
Further processes at NLO QCD accuracy will be included and the BLHA
interface will be extended to include other classes of processes.
\section*{Acknowledgments}
We are grateful to
Ken Arnold, %
Manuel B\"ahr, %
Johannes Bellm, %
Guiseppe Bozzi, %
Martin Brieg, %
Christoph Englert, %
Bastian Feigl, %
Jessica Frank, %
Florian Geyer, %
Nicolas Greiner, %
Christoph Hackstein, %
Vera Hankele, %
Barbara J\"ager, %
Nicolas Kaiser, %
Gunnar Kl\"amke, %
Carlo Oleari, %
Sophy Palmer, %
Stefan Prestel, %
Heidi Rzehak, %
Franziska Schissler, %
Oliver Schlimpert, %
Michael Spannowsky %
and Malgorzata Worek %
for their past contributions to the \Vbfnlo{}\  code.
We also gratefully acknowledge the collaboration of Stefan Kallweit
and Georg Weiglein in the calculation of radiative corrections for specific processes. 

F.C.\ and I.R.\ acknowledge financial support by the AEI-MICINN from
the Spanish Government, and NextGenerationEU funds from the European
Union (Grants No PID2020-113334GB-I00 and CNS2022-136165).  D.N.L.\ is
funded by Phenikaa University under grant number PU2023-1-A-18.
H.S.-S.\ and D.Z.\ were supported in part by the DFG Collaborative
Research Center TRR 257 ``Particle Physics Phenomenology after the
Higgs Discovery''.  M.L.\ acknowledges the support of the Deutsche
Forschungsgemeinschaft (DFG, German Research Association) under
Germany's Excellence Strategy-EXC 2121 ``Quantum Universe''-390833306.

%--------------------------------------------------------------------------------
%================= Section ======================================================
%--------------------------------------------------------------------------------

\bibliography{bib}

\newpage

\appendix

%--------------------------------------------------------------------------------
%================= Section ======================================================
%--------------------------------------------------------------------------------
\onecolumn
\section{Process list}
\label{app:proc_list}
The following is a complete list of all processes available in \Vbfnlo{},
including any Beyond the Standard Model (BSM) effects that are implemented. 
{
\footnotesize
\setlength\LTleft{0pt plus \textwidth minus \textwidth}
\setlength\LTright{0pt plus \textwidth minus \textwidth}
\begin{longtable}{clccccccccc}
\textsc{ProcId} & \textsc{Process} & \rot{\textsc{BLHA}} & \rot{semi-leptonic decay} & \rot{VBF process} & \rot{anom.\ gauge couplings} & \rot{anom.\ Higgs couplings} & \rot{Two-Higgs model} & \rot{Kaluza-Klein model} & \rot{Spin-2 model} & \rot{MSSM} \\
&\\
\hline
\endhead
&\\*
\bf 100 & $p \overset{\mbox{\tiny{(--)}}}{p} \to H \, jj$ &\bsmoptionsBLHA{BHMF}\\*
\bf 101 & $p \overset{\mbox{\tiny{(--)}}}{p} \to H \, jj\to \gamma\gamma \, jj$ &\bsmoptionsBLHA{HMF}\\*
\bf 102 & $p \overset{\mbox{\tiny{(--)}}}{p} \to H \, jj\to \mu^+\mu^- \, jj$ &\bsmoptionsBLHA{HMF}\\*
\bf 103 & $p \overset{\mbox{\tiny{(--)}}}{p} \to H \, jj\to \tau^+\tau^- \, jj$ &\bsmoptionsBLHA{HMF}\\*
\bf 104 & $p \overset{\mbox{\tiny{(--)}}}{p} \to H \, jj\to b\bar{b} \, jj$ &\bsmoptionsBLHA{HMF}\\*
\bf 105 & $p \overset{\mbox{\tiny{(--)}}}{p} \to H \, jj\to W^{+}W^{-} \, jj\to \ell_{1}^+\nu_{\ell_{1}} \ell_{2}^- \bar{\nu}_{\ell_{2}} \,jj$ &\bsmoptionsBLHA{HMF}\\*
\bf 106 & $p \overset{\mbox{\tiny{(--)}}}{p} \to H \, jj\to ZZ \, jj\to \ell_{1}^+ \ell_{1}^- \ell_{2}^+ \ell_{2}^- \,jj$ &\bsmoptionsBLHA{HMF}\\*
\bf 107 & $p \overset{\mbox{\tiny{(--)}}}{p} \to H \, jj\to ZZ \, jj\to \ell_{1}^+ \ell_{1}^- \nu_{\ell_{2}}  \bar{\nu}_{\ell_{2}} \,jj$ &\bsmoptionsBLHA{HMF}\\*
%&\\
%\hline
%&\\
\bf 108 & $p \overset{\mbox{\tiny{(--)}}}{p} \to H \, jj\to W^{+}W^{-} \, jj\to q\bar{q} \, \ell^- \bar{\nu}_{\ell} \,jj$ &\bsmoptionsBLHA{LHMF}\\*
\bf 109 & $p \overset{\mbox{\tiny{(--)}}}{p} \to H \, jj\to W^{+}W^{-} \, jj\to \ell^+\nu_{\ell} \, q\bar{q}  \,jj$ &\bsmoptionsBLHA{LHMF}\\*
\bf 1010 & $p \overset{\mbox{\tiny{(--)}}}{p} \to H \, jj\to ZZ \, jj\to q\bar{q} \, \ell^+ \ell^- \,jj$ &\bsmoptionsBLHA{LHMF}\\*
&\\*
\hline
&\\*
\bf 110 & $p \overset{\mbox{\tiny{(--)}}}{p} \to H \, jjj$ &\bsmoptionsBLHA{F}\\*
\bf 111 & $p \overset{\mbox{\tiny{(--)}}}{p} \to H \, jjj\to \gamma\gamma \, jjj$ &\bsmoptionsBLHA{F}\\*
\bf 112 & $p \overset{\mbox{\tiny{(--)}}}{p} \to H \, jjj\to \mu^+\mu^- \, jjj$ &\bsmoptionsBLHA{F}\\*
\bf 113 & $p \overset{\mbox{\tiny{(--)}}}{p} \to H \, jjj\to \tau^+\tau^- \, jjj$ &\bsmoptionsBLHA{F}\\*
\bf 114 & $p \overset{\mbox{\tiny{(--)}}}{p} \to H \, jjj\to b\bar{b} \, jjj$ &\bsmoptionsBLHA{F}\\*
\bf 115 & $p \overset{\mbox{\tiny{(--)}}}{p} \to H \, jjj\to W^+W^- \, jjj\to \ell_{1}^{+}\nu_{\ell_{1}} \ell_{2}^- \bar{\nu}_{\ell_{2}} \,jjj$ &\bsmoptionsBLHA{F}\\*
\bf 116 & $p \overset{\mbox{\tiny{(--)}}}{p} \to H \, jjj\to ZZ \, jjj\to \ell_{1}^+ \ell_{1}^- \ell_{2}^+ \ell_{2}^- \,jjj$ &\bsmoptionsBLHA{F}\\*
\bf 117 & $p \overset{\mbox{\tiny{(--)}}}{p} \to H \, jjj\to ZZ \, jjj\to \ell_{1}^+ \ell_{1}^- \nu_{\ell_{2}} \bar{\nu}_{\ell_{2}} \,jjj$ &\bsmoptionsBLHA{F}\\*
&\\*
\hline
&\\*
\bf 120 & $p \overset{\mbox{\tiny{(--)}}}{p} \to Z \, jj \to \ell^{+} \ell^{-} \, jj$ &\bsmoptionsBLHA{BVF}\\*
\bf 121 & $p \overset{\mbox{\tiny{(--)}}}{p} \to Z  \, jj\to \nu_\ell \bar{\nu}_\ell \, jj$ &\bsmoptionsBLHA{BVF}\\*
\bf 130 & $p \overset{\mbox{\tiny{(--)}}}{p} \to W^{+} \,  jj\to \ell^{+} \nu_\ell \, jj$ &\bsmoptionsBLHA{BVF}\\*
\bf 140 & $p \overset{\mbox{\tiny{(--)}}}{p} \to W^{-} \, jj\to \ell^{-} \bar{\nu}_\ell  \, jj$ &\bsmoptionsBLHA{BVF}\\*
\bf 150 & $p \overset{\mbox{\tiny{(--)}}}{p} \to \gamma \, jj$ &\bsmoptionsBLHA{BVF}\\*
&\\*
\hline
&\\*
\bf 191 & $p \overset{\mbox{\tiny{(--)}}}{p} \to S_{2}  \, jj\to \gamma \gamma \, jj$ &\bsmoptionsBLHA{SF}\\*
\bf 195 & $p \overset{\mbox{\tiny{(--)}}}{p} \to S_{2}  \, jj\to W^{+}W^{-} \, jj\to \ell_{1}^+\nu_{\ell_{1}} \ell_{2}^- \bar{\nu}_{\ell_{2}} \,jj$ &\bsmoptionsBLHA{SF}\\*
\bf 196 & $p \overset{\mbox{\tiny{(--)}}}{p} \to S_{2}  \, jj\to ZZ \, jj\to \ell_{1}^+ \ell_{1}^- \ell_{2}^+ \ell_{2}^- \,jj$ &\bsmoptionsBLHA{SF}\\*
\bf 197 & $p \overset{\mbox{\tiny{(--)}}}{p} \to S_{2}  \, jj\to  ZZ \, jj\to \ell_{1}^+ \ell_{1}^- \nu_{\ell_{2}} \bar{\nu}_{\ell_{2}} \,jj$ &\bsmoptionsBLHA{SF}\\*
&\\*
\hline
&\\*
\bf 160 & $p \overset{\mbox{\tiny{(--)}}}{p} \to HH \,  jj$ &\bsmoptionsBLHA{F}\\*
\bf 161 & $p \overset{\mbox{\tiny{(--)}}}{p} \to HH \,  jj \to b\bar{b}\tau^+\tau^- \, jj$ &\bsmoptionsBLHA{F}\\*
\bf 162 & $p \overset{\mbox{\tiny{(--)}}}{p} \to HH \,  jj \to b\bar{b}\gamma\gamma \, jj$ &\bsmoptionsBLHA{F}\\*
&\\*
\hline
&\\*
\bf 2100 & $p \overset{\mbox{\tiny{(--)}}}{p} \to H \gamma \, jj$ &\bsmoptionsBLHA{F}\\*
\bf 2101 & $p \overset{\mbox{\tiny{(--)}}}{p} \to H \gamma \, jj\to \gamma\gamma \gamma \, jj$ &\bsmoptionsBLHA{F}\\*
\bf 2102 & $p \overset{\mbox{\tiny{(--)}}}{p} \to H \gamma \, jj\to \mu^+\mu^- \gamma \, jj$ &\bsmoptionsBLHA{F}\\*
\bf 2103 & $p \overset{\mbox{\tiny{(--)}}}{p} \to H \gamma \, jj\to \tau^+\tau^- \gamma \, jj$ &\bsmoptionsBLHA{F}\\*
\bf 2104 & $p \overset{\mbox{\tiny{(--)}}}{p} \to H \gamma \, jj\to b\bar{b} \gamma \, jj$ &\bsmoptionsBLHA{F}\\*
\bf 2105 & $p \overset{\mbox{\tiny{(--)}}}{p} \to H \gamma \, jj\to W^+W^- \gamma \, jj\to \ell_{1}^+\nu_{\ell_{1}} \ell_{2}^- \bar{\nu}_{\ell_{2}} \gamma \,jj$ &\bsmoptionsBLHA{F}\\*
\bf 2106 & $p \overset{\mbox{\tiny{(--)}}}{p} \to H \gamma \, jj\to ZZ \gamma \, jj\to \ell_{1}^+ \ell_{1}^- \ell_{2}^+ \ell_{2}^- \gamma \,jj$ &\bsmoptionsBLHA{F}\\*
\bf 2107 & $p \overset{\mbox{\tiny{(--)}}}{p} \to H \gamma \, jj\to ZZ \gamma \, jj\to \ell_{1}^+ \ell_{1}^- \nu_{\ell_{2}}  \bar{\nu}_{\ell_{2}} \gamma \,jj$ &\bsmoptionsBLHA{F}\\*
&\\*
\hline
&\\*
\bf 200 & $p \overset{\mbox{\tiny{(--)}}}{p} \to W^{+}W^{-} \, jj \to \ell_{1}^{+} \nu_{\ell_{1}} \ell_{2}^{-} \bar{\nu}_{\ell_{2}} \, jj$ &\bsmoptionsBLHA{BVTKSF}\\*
\bf 201 & $p \overset{\mbox{\tiny{(--)}}}{p} \to W^{+}W^{-} \, jj \to q \bar{q} \, \ell^{-} \bar{\nu}_{\ell} \, jj$ &\bsmoptionsBLHA{VTLF}\\*
\bf 202 & $p \overset{\mbox{\tiny{(--)}}}{p} \to W^{+}W^{-}  \, jj\to \ell^{+} \nu_\ell \, q \bar{q} \, jj$ &\bsmoptionsBLHA{VTLF}\\*
\bf 210 & $p \overset{\mbox{\tiny{(--)}}}{p} \to ZZ  \, jj\to \ell_{1}^{+} \ell_{1}^{-} \ell_{2}^{+} \ell_{2}^{-} \, jj$ &\bsmoptionsBLHA{BVTKSF}\\*
\bf 211 & $p \overset{\mbox{\tiny{(--)}}}{p} \to ZZ  \, jj\to \ell_{1}^{+} \ell_{1}^{-} \nu_{\ell_{2}} \bar{\nu}_{\ell_{2}} \, jj$ &\bsmoptionsBLHA{BVTKSF}\\*
\bf 212 & $p \overset{\mbox{\tiny{(--)}}}{p} \to ZZ  \, jj\to q \bar{q} \, \ell^{+} \ell^{-} \, jj$ &\bsmoptionsBLHA{VTLF}\\*
\bf 220 & $p \overset{\mbox{\tiny{(--)}}}{p} \to W^{+}Z \,  jj\to \ell_{1}^{+} \nu_{\ell_{1}} \ell_{2}^{+} \ell_{2}^{-} \, jj$ &\bsmoptionsBLHA{BVTKSF}\\*
\bf 221 & $p \overset{\mbox{\tiny{(--)}}}{p} \to W^{+}Z \,  jj\to q \bar{q} \, \ell^{+} \ell^{-} \, jj$ &\bsmoptionsBLHA{VTLF}\\*
\bf 222 & $p \overset{\mbox{\tiny{(--)}}}{p} \to W^{+}Z \,  jj\to \ell^{+} \nu_{\ell} \, q \bar{q} \, jj$ &\bsmoptionsBLHA{VTLF}\\*
\bf 230 & $p \overset{\mbox{\tiny{(--)}}}{p} \to W^{-}Z \, jj\to \ell_{1}^{-} \bar{\nu}_{\ell _{1}} \ell_{2}^{+} \ell_{2}^{-} \, jj$ &\bsmoptionsBLHA{BVTKSF}\\*
\bf 231 & $p \overset{\mbox{\tiny{(--)}}}{p} \to W^{-}Z \, jj\to q \bar{q} \, \ell^{+} \ell^{-} \, jj$ &\bsmoptionsBLHA{VTLF}\\*
\bf 232 & $p \overset{\mbox{\tiny{(--)}}}{p} \to W^{-}Z \, jj\to \ell^{-} \bar{\nu}_{\ell} \, q \bar{q} \, jj$ &\bsmoptionsBLHA{VTLF}\\*
\bf 240 & $p \overset{\mbox{\tiny{(--)}}}{p} \to \gamma\gamma \, jj$ &\bsmoptionsBLHA{BVF}\\*
\bf 250 & $p \overset{\mbox{\tiny{(--)}}}{p} \to W^{+}W^{+} \,  jj\to \ell_{1}^{+} \nu_{\ell_{1}} \ell_{2}^{+} \nu_{\ell_{2}} \, jj$ &\bsmoptionsBLHA{BVTF}\\*
\bf 251 & $p \overset{\mbox{\tiny{(--)}}}{p} \to W^{+}W^{+} \,  jj\to q \bar{q} \, \ell^{+} \nu_{\ell} \, jj$ &\bsmoptionsBLHA{VTLF}\\*
\bf 260 & $p \overset{\mbox{\tiny{(--)}}}{p} \to W^{-}W^{-} \,  jj\to \ell_{1}^{-} \bar{\nu}_{\ell_{1}} \ell_{2}^{-} \bar{\nu}_{\ell_{2}} \, jj$ &\bsmoptionsBLHA{BVTF}\\*
\bf 261 & $p \overset{\mbox{\tiny{(--)}}}{p} \to W^{-}W^{-} \,  jj\to q \bar{q} \, \ell^{-} \bar{\nu}_{\ell} \, jj$ &\bsmoptionsBLHA{VTLF}\\*
\bf 270 & $p \overset{\mbox{\tiny{(--)}}}{p} \to W^{+}\gamma \, jj\to \ell^{+} \nu_{\ell} \gamma \, jj$ &\bsmoptionsBLHA{BVF}\\*
\bf 280 & $p \overset{\mbox{\tiny{(--)}}}{p} \to W^{-}\gamma \, jj\to \ell^{-} \bar{\nu}_{\ell} \gamma \, jj$ &\bsmoptionsBLHA{BVF}\\*
\bf 290 & $p \overset{\mbox{\tiny{(--)}}}{p} \to Z\gamma \, jj\to \ell^{+} \ell^{-} \gamma \, jj$ &\bsmoptionsBLHA{BVF}\\*
\bf 291 & $p \overset{\mbox{\tiny{(--)}}}{p} \to Z\gamma \, jj\to \nu_{\ell}\bar{\nu}_{\ell} \gamma \, jj$ &\bsmoptionsBLHA{BVF}\\*
&\\*
\hline
&\\*
\bf 3120 & $p \overset{\mbox{\tiny{(--)}}}{p} \to Z \,  jj\to \ell^{+} \ell^{-} \, jj$  &\bsmoptionsBLHA{}\\*
\bf 3121 & $p \overset{\mbox{\tiny{(--)}}}{p} \to Z \,  jj\to \nu_{\ell} \bar{\nu}_{\ell} \, jj$    &\bsmoptionsBLHA{}\\*
\bf 3130 & $p \overset{\mbox{\tiny{(--)}}}{p} \to W^{+} \,  jj\to \ell^{+} \nu_\ell \, jj$ &\bsmoptionsBLHA{}\\*
\bf 3140 & $p \overset{\mbox{\tiny{(--)}}}{p} \to W^{-} \, jj\to \ell^{-} \bar{\nu}_\ell  \, jj$ &\bsmoptionsBLHA{}\\*
&\\*
\hline
&\\*
\bf 3210 & $p \overset{\mbox{\tiny{(--)}}}{p} \to ZZ \,  jj\to \ell_{1}^{+} \ell_{1}^{-} \ell_{2}^{+} \ell_{2}^{-} \, jj$   &\bsmoptionsBLHA{}\\*
\bf 3211 & $p \overset{\mbox{\tiny{(--)}}}{p} \to ZZ \,  jj\to \ell_{1}^{+} \ell_{1}^{-} \nu_{\ell_{2}} \bar{\nu}_{\ell_{2}} \, jj$  &\bsmoptionsBLHA{}\\*
\bf 3220 & $p \overset{\mbox{\tiny{(--)}}}{p} \to W^{+}Z \,  jj\to \ell_{1}^{+} \nu_{\ell_{1}} \ell_{2}^{+} \ell_{2}^{-} \, jj$ &\bsmoptionsBLHA{}\\*
\bf 3230 & $p \overset{\mbox{\tiny{(--)}}}{p} \to W^{-}Z \, jj\to \ell_{1}^{-} \bar{\nu}_{\ell _{1}} \ell_{2}^{+} \ell_{2}^{-} \, jj$ &\bsmoptionsBLHA{}\\*
\bf 3240 & $p \overset{\mbox{\tiny{(--)}}}{p} \to \gamma\gamma \,  jj$  &\bsmoptionsBLHA{}\\*
\bf 3250 & $p \overset{\mbox{\tiny{(--)}}}{p} \to W^{+}W^{+} \,  jj\to \ell_{1}^{+} \nu_{\ell_{1}} \ell_{2}^{+} \nu_{\ell_{2}} \, jj$ &\bsmoptionsBLHA{}\\*
\bf 3260 & $p \overset{\mbox{\tiny{(--)}}}{p} \to W^{-}W^{-} \,  jj\to \ell_{1}^{-} \bar{\nu}_{\ell_{1}} \ell_{2}^{-} \bar{\nu}_{\ell_{2}} \, jj$ &\bsmoptionsBLHA{}\\*
\bf 3270 & $p \overset{\mbox{\tiny{(--)}}}{p} \to W^{+}\gamma \,  jj\to \ell^{+} \nu_{\ell} \gamma \, jj$ &\bsmoptionsBLHA{}\\*
\bf 3280 & $p \overset{\mbox{\tiny{(--)}}}{p} \to W^{-}\gamma \, jj\to \ell^{-} \bar{\nu}_{\ell} \gamma \, jj$ &\bsmoptionsBLHA{}\\*
\bf 3290 & $p \overset{\mbox{\tiny{(--)}}}{p} \to Z\gamma \,  jj\to \ell^{+} \ell^{-} \gamma \, jj$  &\bsmoptionsBLHA{}\\*
\bf 3291 & $p \overset{\mbox{\tiny{(--)}}}{p} \to Z\gamma \,  jj\to \nu_{\ell} \bar{\nu}_{\ell} \gamma\, jj$    &\bsmoptionsBLHA{}\\*
&\\*
\hline
&\\*
\bf 1330 & $p \overset{\mbox{\tiny{(--)}}}{p} \to W^+ \to \ell^+\nu_{\ell} $ &\bsmoptionsBLHA{}\\*
\bf 1340 & $p \overset{\mbox{\tiny{(--)}}}{p} \to W^- \to \ell^- \bar{\nu}_{\ell} $ &\bsmoptionsBLHA{}\\*
\bf 1630 & $p \overset{\mbox{\tiny{(--)}}}{p} \to W^+ \, j \to \ell^+\nu_{\ell} \, j $ &\bsmoptionsBLHA{}\\*
\bf 1640 & $p \overset{\mbox{\tiny{(--)}}}{p} \to W^- \, j\to \ell^- \bar{\nu}_{\ell} \, j $ &\bsmoptionsBLHA{}\\*
&\\*
\hline
&\\*
\bf 300 & $p \overset{\mbox{\tiny{(--)}}}{p} \to W^{+}W^{-} \to \ell_{1}^{+} \nu_{\ell_{1}} \ell_{2}^{-}\bar{\nu}_{\ell_{2}} $ &\bsmoptionsBLHA{BHV}\\*
\bf 301 & $p \overset{\mbox{\tiny{(--)}}}{p} \to W^{+}W^{-} \to q \bar{q} \, \ell^{-}\bar{\nu}_{\ell} $ &\bsmoptionsBLHA{HVL}\\*
\bf 302 & $p \overset{\mbox{\tiny{(--)}}}{p} \to W^{+}W^{-} \to \ell^{+} \nu_{\ell} \, q \bar{q} $ &\bsmoptionsBLHA{HVL}\\*
\bf 310 & $p \overset{\mbox{\tiny{(--)}}}{p} \to W^{+}Z \to  \ell_{1}^{+} \nu_{\ell_1}  \ell_{2}^{+} \ell_{2}^{-} $ &\bsmoptionsBLHA{BV}\\*
\bf 312 & $p \overset{\mbox{\tiny{(--)}}}{p} \to W^{+}Z \to  q \bar{q} \, \ell^{+} \ell^{-} $ &\bsmoptionsBLHA{VL}\\*
\bf 313 & $p \overset{\mbox{\tiny{(--)}}}{p} \to W^{+}Z \to  \ell^{+} \nu_{\ell} \, q \bar{q} $ &\bsmoptionsBLHA{VL}\\*
\bf 320 & $p \overset{\mbox{\tiny{(--)}}}{p} \to W^{-}Z \to \ell_{1}^{-} \bar{\nu}_{\ell_{1}}  \ell_{2}^{+} \ell_{2}^{-} $ &\bsmoptionsBLHA{BV}\\*
\bf 322 & $p \overset{\mbox{\tiny{(--)}}}{p} \to W^{-}Z \to q \bar{q} \, \ell^{+} \ell^{-} $ &\bsmoptionsBLHA{VL}\\*
\bf 323 & $p \overset{\mbox{\tiny{(--)}}}{p} \to W^{-}Z \to \ell^{-} \bar{\nu}_{\ell} \, q \bar{q} $ &\bsmoptionsBLHA{VL}\\*
\bf 330 & $p \overset{\mbox{\tiny{(--)}}}{p} \to ZZ \to \ell_{1}^{-} \ell_{1}^{+}  \ell_{2}^{-} \ell_{2}^{+} $ &\bsmoptionsBLHA{BH}\\*
\bf 331 & $p \overset{\mbox{\tiny{(--)}}}{p} \to ZZ \to q \bar{q} \, \ell^{-} \ell^{+} $ &\bsmoptionsBLHA{HL}\\*
\bf 340 & $p \overset{\mbox{\tiny{(--)}}}{p} \to W^{+}\gamma \to \ell_{1}^{+} \nu_{\ell_1} \gamma $ &\bsmoptionsBLHA{BV}\\*
\bf 350 & $p \overset{\mbox{\tiny{(--)}}}{p} \to W^{-}\gamma \to \ell_{1}^{-} \bar{\nu}_{\ell_{1}} \gamma $ &\bsmoptionsBLHA{BV}\\*
\bf 360 & $p \overset{\mbox{\tiny{(--)}}}{p} \to Z\gamma \to \ell_{1}^{-} \ell_{1}^{+}  \gamma $ &\bsmoptionsBLHA{BH}\\*
\bf 370 & $p \overset{\mbox{\tiny{(--)}}}{p} \to \gamma\gamma $ &\bsmoptionsBLHA{BH}\\*
&\\*
\hline
&\\*
\bf 1300 & $p \overset{\mbox{\tiny{(--)}}}{p} \to W^+ H \to \ell^+\nu_{\ell} H $ &\bsmoptionsBLHA{V}\\*
\bf 1301 & $p \overset{\mbox{\tiny{(--)}}}{p} \to W^+ H \to \ell^+\nu_{\ell} \gamma\gamma $ &\bsmoptionsBLHA{V}\\*
\bf 1302 & $p \overset{\mbox{\tiny{(--)}}}{p} \to W^+ H \to \ell^+\nu_{\ell} \mu^+\mu^- $ &\bsmoptionsBLHA{V}\\*
\bf 1303 & $p \overset{\mbox{\tiny{(--)}}}{p} \to W^+ H \to \ell^+\nu_{\ell} \tau^+\tau^- $ &\bsmoptionsBLHA{V}\\*
\bf 1304 & $p \overset{\mbox{\tiny{(--)}}}{p} \to W^+ H \to \ell^+\nu_{\ell} b\bar{b} $ &\bsmoptionsBLHA{V}\\*
\bf 1305 & $p \overset{\mbox{\tiny{(--)}}}{p} \to W^+ H \to W^+ W^{+}W^{-} \to \ell_{1}^+\nu_{\ell_{1}} \ell_{2}^+\nu_{\ell_{2}} \ell_{3}^- \bar{\nu}_{\ell_{3}}$ &\bsmoptionsBLHA{V}\\*
\bf 1306 & $p \overset{\mbox{\tiny{(--)}}}{p} \to W^+ H \to W^+ ZZ \to \ell_{1}^+\nu_{\ell_{1}} \ell_{2}^+ \ell_{2}^- \ell_{3}^+ \ell_{3}^-$ &\bsmoptionsBLHA{V}\\*
\bf 1307 & $p \overset{\mbox{\tiny{(--)}}}{p} \to W^+ H \to W^+ ZZ \to \ell_{1}^+\nu_{\ell_{1}} \ell_{2}^+ \ell_{2}^- \nu_{\ell_{3}}  \bar{\nu}_{\ell_{3}}$ &\bsmoptionsBLHA{V}\\*
&\\*
\hline
&\\*
\bf 1310 & $p \overset{\mbox{\tiny{(--)}}}{p} \to W^- H \to \ell^- \bar{\nu}_{\ell} H $ &\bsmoptionsBLHA{V}\\*
\bf 1311 & $p \overset{\mbox{\tiny{(--)}}}{p} \to W^- H \to \ell^- \bar{\nu}_{\ell} \gamma\gamma $ &\bsmoptionsBLHA{V}\\*
\bf 1312 & $p \overset{\mbox{\tiny{(--)}}}{p} \to W^- H \to \ell^- \bar{\nu}_{\ell} \mu^+\mu^- $ &\bsmoptionsBLHA{V}\\*
\bf 1313 & $p \overset{\mbox{\tiny{(--)}}}{p} \to W^- H \to \ell^- \bar{\nu}_{\ell} \tau^+\tau^- $ &\bsmoptionsBLHA{V}\\*
\bf 1314 & $p \overset{\mbox{\tiny{(--)}}}{p} \to W^- H \to \ell^- \bar{\nu}_{\ell} b\bar{b} $ &\bsmoptionsBLHA{V}\\*
\bf 1315 & $p \overset{\mbox{\tiny{(--)}}}{p} \to W^- H \to W^- W^{+}W^{-} \to \ell_{1}^-\bar{\nu}_{\ell_{1}} \ell_{2}^+\nu_{\ell_{2}} \ell_{3}^- \bar{\nu}_{\ell_{3}}$ &\bsmoptionsBLHA{V}\\*
\bf 1316 & $p \overset{\mbox{\tiny{(--)}}}{p} \to W^- H \to W^- ZZ \to \ell_{1}^-\bar{\nu}_{\ell_{1}} \ell_{2}^+ \ell_{2}^- \ell_{3}^+ \ell_{3}^-$ &\bsmoptionsBLHA{V}\\*
\bf 1317 & $p \overset{\mbox{\tiny{(--)}}}{p} \to W^- H \to W^- ZZ \to \ell_{1}^-\bar{\nu}_{\ell_{1}} \ell_{2}^+ \ell_{2}^- \nu_{\ell_{3}}  \bar{\nu}_{\ell_{3}}$ &\bsmoptionsBLHA{V}\\*
&\\*
\hline
&\\*
\bf 600 & $p \overset{\mbox{\tiny{(--)}}}{p}  \to W^{+} W^{-} j \to \ell_{1}^{+}\nu_{\ell_1} \ell_{2}^{-} \bar{\nu}_{\ell_2} j $ & \bsmoptionsBLHA{H}\\*
\bf 601 & $p \overset{\mbox{\tiny{(--)}}}{p}  \to W^{+} W^{-} j \to q\bar{q} \ell^{-} \bar{\nu}_{\ell} j $ & \bsmoptionsBLHA{L}\\*
\bf 602 & $p \overset{\mbox{\tiny{(--)}}}{p}  \to W^{+} W^{-} j \to \ell^{+}\nu_{\ell} q\bar{q} j $ & \bsmoptionsBLHA{L}\\*
\bf 610 & $p \overset{\mbox{\tiny{(--)}}}{p}  \to W^{-} \gamma j \to \ell^{-} \bar \nu_{\ell} \gamma j $ &\bsmoptionsBLHA{V}\\*
\bf 620 & $p \overset{\mbox{\tiny{(--)}}}{p}  \to W^{+} \gamma j  \to \ell^{+} \nu_{\ell} \gamma j $ &\bsmoptionsBLHA{V}\\*
\bf 630 & $p \overset{\mbox{\tiny{(--)}}}{p}  \to W^{-} Z j \to \ell_{1}^{-} \bar \nu_{\ell_1} \ell_{2}^{-} \ell_{2}^{+} j$ &\bsmoptionsBLHA{V}\\*
\bf 640 & $p \overset{\mbox{\tiny{(--)}}}{p}  \to W^{+} Z j \to  \ell_{1}^{+}\nu_{\ell_1} \ell_{2}^{-} \ell_{2}^{+}j $ &\bsmoptionsBLHA{V}\\*
\bf 650 & $p \overset{\mbox{\tiny{(--)}}}{p}  \to Z Z j \to \ell_{1}^{+} \ell_{1}^{-} \ell_{2}^{+} \ell_{2}^{-} j $ & \bsmoptionsBLHA{H}\\*
&\\*
\hline
&\\*
\bf 1600 & $p \overset{\mbox{\tiny{(--)}}}{p} \to W^+ H \, j \to \ell^+\nu_{\ell} H \, j $ &\bsmoptionsBLHA{V}\\*
\bf 1601 & $p \overset{\mbox{\tiny{(--)}}}{p} \to W^+ H \, j \to \ell^+\nu_{\ell} \gamma\gamma \, j $ &\bsmoptionsBLHA{V}\\*
\bf 1602 & $p \overset{\mbox{\tiny{(--)}}}{p} \to W^+ H \, j \to \ell^+\nu_{\ell} \mu^+\mu^- \, j $ &\bsmoptionsBLHA{V}\\*
\bf 1603 & $p \overset{\mbox{\tiny{(--)}}}{p} \to W^+ H \, j \to \ell^+\nu_{\ell} \tau^+\tau^- \, j $ &\bsmoptionsBLHA{V}\\*
\bf 1604 & $p \overset{\mbox{\tiny{(--)}}}{p} \to W^+ H \, j \to \ell^+\nu_{\ell} b\bar{b} \, j $ &\bsmoptionsBLHA{V}\\*
\bf 1605 & $p \overset{\mbox{\tiny{(--)}}}{p} \to W^+ H \, j \to W^+ W^{+}W^{-} \, j \to \ell_{1}^+\nu_{\ell_{1}} \ell_{2}^+\nu_{\ell_{2}} \ell_{3}^- \bar{\nu}_{\ell_{3}}\, j $ &\bsmoptionsBLHA{V}\\*
\bf 1606 & $p \overset{\mbox{\tiny{(--)}}}{p} \to W^+ H \, j \to W^+ ZZ \, j \to \ell_{1}^+\nu_{\ell_{1}} \ell_{2}^+ \ell_{2}^- \ell_{3}^+ \ell_{3}^-\, j $ &\bsmoptionsBLHA{V}\\*
\bf 1607 & $p \overset{\mbox{\tiny{(--)}}}{p} \to W^+ H \, j \to W^+ ZZ \, j \to \ell_{1}^+\nu_{\ell_{1}} \ell_{2}^+ \ell_{2}^- \nu_{\ell_{3}}  \bar{\nu}_{\ell_{3}}\, j $ &\bsmoptionsBLHA{V}\\*
&\\*
\hline
&\\*
\bf 1610 & $p \overset{\mbox{\tiny{(--)}}}{p} \to W^- H \, j \to \ell^- \bar{\nu}_{\ell} H \, j $ &\bsmoptionsBLHA{V}\\*
\bf 1611 & $p \overset{\mbox{\tiny{(--)}}}{p} \to W^- H \, j \to \ell^- \bar{\nu}_{\ell} \gamma\gamma \, j $ &\bsmoptionsBLHA{V}\\*
\bf 1612 & $p \overset{\mbox{\tiny{(--)}}}{p} \to W^- H \, j \to \ell^- \bar{\nu}_{\ell} \mu^+\mu^- \, j $ &\bsmoptionsBLHA{V}\\*
\bf 1613 & $p \overset{\mbox{\tiny{(--)}}}{p} \to W^- H \, j \to \ell^- \bar{\nu}_{\ell} \tau^+\tau^- \, j $ &\bsmoptionsBLHA{V}\\*
\bf 1614 & $p \overset{\mbox{\tiny{(--)}}}{p} \to W^- H \, j \to \ell^- \bar{\nu}_{\ell} b\bar{b} \, j $ &\bsmoptionsBLHA{V}\\*
\bf 1615 & $p \overset{\mbox{\tiny{(--)}}}{p} \to W^- H \, j \to W^- W^{+}W^{-} \, j \to \ell_{1}^-\bar{\nu}_{\ell_{1}} \ell_{2}^+\nu_{\ell_{2}} \ell_{3}^- \bar{\nu}_{\ell_{3}}\, j $ &\bsmoptionsBLHA{V}\\*
\bf 1616 & $p \overset{\mbox{\tiny{(--)}}}{p} \to W^- H \, j \to W^- ZZ \, j \to \ell_{1}^-\bar{\nu}_{\ell_{1}} \ell_{2}^+ \ell_{2}^- \ell_{3}^+ \ell_{3}^-\, j $ &\bsmoptionsBLHA{V}\\*
\bf 1617 & $p \overset{\mbox{\tiny{(--)}}}{p} \to W^- H \, j \to W^- ZZ \, j \to \ell_{1}^-\bar{\nu}_{\ell_{1}} \ell_{2}^+ \ell_{2}^- \nu_{\ell_{3}}  \bar{\nu}_{\ell_{3}}\, j $ &\bsmoptionsBLHA{V}\\*
&\\*
\hline
&\\*
\bf 400 & $p \overset{\mbox{\tiny{(--)}}}{p} \to W^{+}W^{-}Z \to \ell_{1}^{+}\nu_{\ell_{1}} \ell_{2}^{-} \bar{\nu}_{\ell_{2}} \ell_{3}^{+} \ell_{3}^{-} $ &\bsmoptionsBLHA{BVK}\\*
\bf 401 & $p \overset{\mbox{\tiny{(--)}}}{p} \to W^{+}W^{-}Z \to q \bar{q} \, \ell_{1}^{-} \bar{\nu}_{\ell_{1}} \ell_{2}^{+} \ell_{2}^{-} $ &\bsmoptionsBLHA{LV}\\*
\bf 402 & $p \overset{\mbox{\tiny{(--)}}}{p} \to W^{+}W^{-}Z \to \ell_{1}^{+}\nu_{\ell_{1}} \, q \bar{q} \, \ell_{2}^{+} \ell_{2}^{-} $ &\bsmoptionsBLHA{LV}\\*
\bf 403 & $p \overset{\mbox{\tiny{(--)}}}{p} \to W^{+}W^{-}Z \to \ell_{1}^{+}\nu_{\ell_{1}} \ell_{2}^{-} \bar{\nu}_{\ell_{2}} \, q \bar{q} $ &\bsmoptionsBLHA{LV}\\*
\bf 410 & $p \overset{\mbox{\tiny{(--)}}}{p} \to ZZW^{+} \to  \ell_{1}^{+} \ell_{1}^{-}  \ell_{2}^{+} \ell_{2}^{-} \ell_{3}^{+} \nu_{\ell_{3}} $ &\bsmoptionsBLHA{BVK}\\*
\bf 411 & $p \overset{\mbox{\tiny{(--)}}}{p} \to ZZW^{+} \to  \ell_{1}^{+} \ell_{1}^{-}  \ell_{2}^{+} \ell_{2}^{-} \, q \bar{q} $ &\bsmoptionsBLHA{LV}\\*
\bf 412 & $p \overset{\mbox{\tiny{(--)}}}{p} \to ZZW^{+} \to  q \bar{q} \, \ell_{1}^{+} \ell_{1}^{-} \ell_{2}^{+} \nu_{\ell_{2}} $ &\bsmoptionsBLHA{LV}\\*
\bf 420 & $p \overset{\mbox{\tiny{(--)}}}{p} \to ZZW^{-} \to \ell_{1}^{+} \ell_{1}^{-}  \ell_{2}^{+} \ell_{2}^{-} \ell_{3}^{-}  \bar{\nu}_{\ell_{3}}$ &\bsmoptionsBLHA{BVK}\\*
\bf 421 & $p \overset{\mbox{\tiny{(--)}}}{p} \to ZZW^{-} \to \ell_{1}^{+} \ell_{1}^{-}  \ell_{2}^{+} \ell_{2}^{-} \, q \bar{q} $ &\bsmoptionsBLHA{LV}\\*
\bf 422 & $p \overset{\mbox{\tiny{(--)}}}{p} \to ZZW^{-} \to q \bar{q} \, \ell_{1}^{+} \ell_{1}^{-} \ell_{2}^{-}  \bar{\nu}_{\ell_{2}}$ &\bsmoptionsBLHA{LV}\\*
\bf 430 & $p \overset{\mbox{\tiny{(--)}}}{p} \to W^{+}W^{-}W^{+} \to \ell_{1}^{+}\nu_{\ell_1} \ell_{2}^{-} \bar{\nu}_{\ell_2} \ell_{3}^{+}\nu_{\ell_{3}}$ &\bsmoptionsBLHA{BVK}\\*
\bf 431 & $p \overset{\mbox{\tiny{(--)}}}{p} \to W^{+}W^{-}W^{+} \to  q \bar{q} \, \ell_{1}^{-} \bar{\nu}_{\ell_1} \ell_{2}^{+}\nu_{\ell_{2}}$ &\bsmoptionsBLHA{LV}\\*
\bf 432 & $p \overset{\mbox{\tiny{(--)}}}{p} \to W^{+}W^{-}W^{+} \to \ell_{1}^{+}\nu_{\ell_1} \, q \bar{q} \, \ell_{2}^{+}\nu_{\ell_{2}}$ &\bsmoptionsBLHA{LV}\\*
\bf 440 & $p \overset{\mbox{\tiny{(--)}}}{p} \to W^{-}W^{+}W^{-} \to \ell_{1}^{-} \bar{\nu}_{\ell_1}\ell_{2}^{+}\nu_{\ell_2} \ell_{3}^{-} \bar{\nu}_{\ell_{3}} $ &\bsmoptionsBLHA{BVK}\\*
\bf 441 & $p \overset{\mbox{\tiny{(--)}}}{p} \to W^{-}W^{+}W^{-} \to \ell_{1}^{-} \bar{\nu}_{\ell_1} \, q \bar{q} \, \ell_{2}^{-} \bar{\nu}_{\ell_{2}} $ &\bsmoptionsBLHA{LV}\\*
\bf 442 & $p \overset{\mbox{\tiny{(--)}}}{p} \to W^{-}W^{+}W^{-} \to  q \bar{q} \, \ell_{1}^{+}\nu_{\ell_1} \ell_{2}^{-} \bar{\nu}_{\ell_{2}} $ &\bsmoptionsBLHA{LV}\\*
\bf 450 & $p \overset{\mbox{\tiny{(--)}}}{p} \to ZZZ \to \ell_{1}^{-} \ell_{1}^{+} \ell_{2}^{-} \ell_{2}^{+} \ell_{3}^{-} \ell_{3}^{+} $ &\bsmoptionsBLHA{BV}\\*
\bf 451 & $p \overset{\mbox{\tiny{(--)}}}{p} \to ZZZ \to  q \bar{q} \, \ell_{1}^{-} \ell_{1}^{+} \ell_{2}^{-} \ell_{2}^{+}  $ &\bsmoptionsBLHA{LV}\\*
&\\*
\hline
&\\*
\bf 460 & $p \overset{\mbox{\tiny{(--)}}}{p} \to W^{-}W^{+} \gamma \to \ell_{1}^{-} \bar{\nu}_{\ell_1} \ell_{2}^{+}\nu_{\ell_2} \gamma$ &\bsmoptionsBLHA{BV}\\*
\bf 461 & $p \overset{\mbox{\tiny{(--)}}}{p} \to W^{+}W^{-} \gamma \to  q \bar{q} \, \ell^{-}\bar{\nu}_{\ell} \gamma$ &\bsmoptionsBLHA{LV}\\*
\bf 462 & $p \overset{\mbox{\tiny{(--)}}}{p} \to W^{+}W^{-} \gamma \to \ell^{+} \nu_{\ell} \, q \bar{q} \, \gamma$ &\bsmoptionsBLHA{LV}\\*
\bf 470 & $p \overset{\mbox{\tiny{(--)}}}{p} \to Z Z \gamma \to \ell_{1}^{-} \ell_{1}^{+} \ell_{2}^{-} \ell_{2}^{+} \gamma$ &\bsmoptionsBLHA{BV}\\*
\bf 471 & $p \overset{\mbox{\tiny{(--)}}}{p} \to Z Z \gamma \to \ell^{-} \ell^{+} \, q \bar{q} \, \gamma$ &\bsmoptionsBLHA{LV}\\*
\bf 472 & $p \overset{\mbox{\tiny{(--)}}}{p} \to Z Z \gamma \to \ell_{1}^{-} \ell_{1}^{+} \nu_{\ell_2} \bar{\nu}_{\ell_2} \gamma$ &\bsmoptionsBLHA{B}\\*
\bf 480 & $p \overset{\mbox{\tiny{(--)}}}{p} \to W^{+} Z \gamma \to \ell_{1}^{+}\nu_{\ell_1} \ell_{2}^{-} \ell_{2}^{+} \gamma$ &\bsmoptionsBLHA{BV}\\*
\bf 481 & $p \overset{\mbox{\tiny{(--)}}}{p} \to W^{+} Z \gamma \to  q \bar{q} \, \ell^{-} \ell^{+} \gamma$ &\bsmoptionsBLHA{LV}\\*
\bf 482 & $p \overset{\mbox{\tiny{(--)}}}{p} \to W^{+} Z \gamma \to \ell^{+}\nu_{\ell} \, q \bar{q} \, \gamma$ &\bsmoptionsBLHA{LV}\\*
\bf 490 & $p \overset{\mbox{\tiny{(--)}}}{p} \to W^{-} Z \gamma \to \ell_{1}^{-} \bar{\nu}_{\ell_1} \ell_{2}^{-} \ell_{2}^{+} \gamma$ &\bsmoptionsBLHA{BV}\\*
\bf 491 & $p \overset{\mbox{\tiny{(--)}}}{p} \to W^{-} Z \gamma \to  q \bar{q} \, \ell^{-} \ell^{+} \gamma$ &\bsmoptionsBLHA{LV}\\*
\bf 492 & $p \overset{\mbox{\tiny{(--)}}}{p} \to W^{-} Z \gamma \to \ell^{-} \bar{\nu}_{\ell} \, q \bar{q} \, \gamma$ &\bsmoptionsBLHA{LV}\\*
\bf 500 & $p \overset{\mbox{\tiny{(--)}}}{p} \to W^{+} \gamma \gamma \to {\ell}^{+}\nu_{\ell} \gamma \gamma$ &\bsmoptionsBLHA{BV}\\*
\bf 510 & $p \overset{\mbox{\tiny{(--)}}}{p} \to W^{-} \gamma \gamma \to {\ell}^{-} \bar{\nu}_{\ell} \gamma \gamma$ &\bsmoptionsBLHA{BV}\\*
\bf 520 & $p \overset{\mbox{\tiny{(--)}}}{p} \to Z \gamma \gamma \to {\ell}^{-} {\ell}^{+} \gamma \gamma$ &\bsmoptionsBLHA{BV}\\*
\bf 521 & $p \overset{\mbox{\tiny{(--)}}}{p} \to Z \gamma \gamma \to \nu_{\ell} \bar{\nu}_{\ell} \gamma \gamma$ &\bsmoptionsBLHA{BV}\\*
\bf 530 & $p \overset{\mbox{\tiny{(--)}}}{p} \to \gamma \gamma \gamma $ &\bsmoptionsBLHA{B}\\*
&\\*
\hline
&\\*
\bf 800 & $p \overset{\mbox{\tiny{(--)}}}{p}  \to W^{+} \gamma \gamma j  \to \ell^{+} \nu_{\ell} \gamma \gamma j $ &\bsmoptionsBLHA{V}\\*
\bf 810 & $p \overset{\mbox{\tiny{(--)}}}{p}  \to W^{-} \gamma \gamma j \to \ell^{-} \bar \nu_{\ell} \gamma \gamma j $ &\bsmoptionsBLHA{V}\\*
&\\*
\hline
\end{longtable}
}

\clearpage
The gluon-fusion processes are given below.
{
\footnotesize
\setlength\LTleft{0pt plus \textwidth minus \textwidth}
\setlength\LTright{0pt plus \textwidth minus \textwidth}
\begin{longtable}{clccccccccc}
\textsc{ProcId} & \textsc{Process} &\rot{BLHA} &  \rot{gluon-fusion process} & \rot{semi-leptonic decay} & \rot{anom.\ Higgs couplings} & \rot{general 2HDM} & \rot{MSSM}\\
&\\
\hline
\endhead
&\\*
\bf 4100 & $p \overset{\mbox{\tiny{(--)}}}{p} \to H \, jj $ &\bsmgfoptionsBLHA{GTM}\\*
\bf 4101 & $p \overset{\mbox{\tiny{(--)}}}{p} \to H \, jj\to \gamma\gamma \, jj$ &\bsmgfoptionsBLHA{GM}\\*
\bf 4102 & $p \overset{\mbox{\tiny{(--)}}}{p} \to H \, jj\to \mu^+\mu^- \, jj$ &\bsmgfoptionsBLHA{GM}\\*
\bf 4103 & $p \overset{\mbox{\tiny{(--)}}}{p} \to H \, jj\to \tau^+\tau^- \, jj$ &\bsmgfoptionsBLHA{GM}\\*
\bf 4104 & $p \overset{\mbox{\tiny{(--)}}}{p} \to H \, jj\to b\bar{b} \, jj$ &\bsmgfoptionsBLHA{GM}\\*
\bf 4105 & $p \overset{\mbox{\tiny{(--)}}}{p} \to H \, jj\to W^{+}W^{-} \, jj\to \ell_{1}^+\nu_{\ell_{1}} \ell_{2}^- \bar{\nu}_{\ell_{2}} \,jj$ &\bsmgfoptionsBLHA{GHTM}\\*
\bf 4106 & $p \overset{\mbox{\tiny{(--)}}}{p} \to H \, jj\to ZZ \, jj\to \ell_{1}^+ \ell_{1}^- \ell_{2}^+ \ell_{2}^- \,jj$ &\bsmgfoptionsBLHA{GHTM}\\*
\bf 4107 & $p \overset{\mbox{\tiny{(--)}}}{p} \to H \, jj\to ZZ \, jj\to \ell_{1}^+ \ell_{1}^- \nu_{\ell_{2}}  \bar{\nu}_{\ell_{2}} \,jj$ &\bsmgfoptionsBLHA{GHTM}\\*
&\\*
\hline
\bf 4200 & $p \overset{\mbox{\tiny{(--)}}}{p} \to H \, jjj $ &\bsmgfoptionsBLHA{GTM}\\*
\hline
&\\*
\bf 4300 & $p \overset{\mbox{\tiny{(--)}}}{p} \to W^{+}W^{-} \to \ell_{1}^{+} \nu_{\ell_{1}} \ell_{2}^{-}\bar{\nu}_{\ell_{2}} $ &\bsmgfoptionsBLHA{GH}\\*
\bf 4301 & $p \overset{\mbox{\tiny{(--)}}}{p} \to W^{+}W^{-} \to q \bar{q} \, \ell^{-}\bar{\nu}_{\ell} $ &\bsmgfoptionsBLHA{GHL}\\*
\bf 4302 & $p \overset{\mbox{\tiny{(--)}}}{p} \to W^{+}W^{-} \to \ell^{+} \nu_{\ell} \, q \bar{q} $ &\bsmgfoptionsBLHA{GHL}\\*
\bf 4330 & $p \overset{\mbox{\tiny{(--)}}}{p} \to ZZ \to \ell_{1}^{-} \ell_{1}^{+}  \ell_{2}^{-} \ell_{2}^{+} $ &\bsmgfoptionsBLHA{GH}\\*
\bf 4331 & $p \overset{\mbox{\tiny{(--)}}}{p} \to ZZ \to q \bar{q} \, \ell^{-} \ell^{+} $ &\bsmgfoptionsBLHA{GHL}\\*
\bf 4360 & $p \overset{\mbox{\tiny{(--)}}}{p} \to Z\gamma \to \ell_{1}^{-} \ell_{1}^{+}  \gamma $ &\bsmgfoptionsBLHA{GH}\\*
\bf 4370 & $p \overset{\mbox{\tiny{(--)}}}{p} \to \gamma\gamma $ &\bsmgfoptionsBLHA{GH}\\*
&\\*
\hline
&\\*
\bf 4600 & $p \overset{\mbox{\tiny{(--)}}}{p} \to W^{+}W^{-} j \to \ell_{1}^{+} \nu_{\ell_{1}} \ell_{2}^{-}\bar{\nu}_{\ell_{2}} j $ & \bsmgfoptionsBLHA{GH}\\*
\bf 4601 & $p \overset{\mbox{\tiny{(--)}}}{p} \to W^{+}W^{-} j \to q\bar{q} \ell^{-}\bar{\nu}_{\ell} j $ & \bsmgfoptionsBLHA{GL}\\*
\bf 4602 & $p \overset{\mbox{\tiny{(--)}}}{p} \to W^{+}W^{-} j \to \ell^{+} \nu_{\ell} q\bar{q} j $ & \bsmgfoptionsBLHA{GL}\\*
\bf 4650 & $p \overset{\mbox{\tiny{(--)}}}{p} \to ZZ j \to \ell_{1}^{-} \ell_{1}^{+}  \ell_{2}^{-} \ell_{2}^{+} j $ & \bsmgfoptionsBLHA{GH}\\*
&\\* 
\hline
\end{longtable}
}

\twocolumn

%% end contents %%%%%%%%%%%%%%%%%%%%%%%%%%%%%%%%%%%%%%%%%%%%%%%%%%%%%%%%%%%%%%%%
%%%%%%%%%%%%%%%%%%%%%%%%%%%%%%%%%%%%%%%%%%%%%%%%%%%%%%%%%%%%%%%%%%%%%%%%%%%%%%%%
\end{document}